# Negative ion formation in lanthanide atoms: Many-body effects

Zineb Felfli[1,*], Alfred Z. Msezane[1] and Dmitri Sokolovski[2]

[1]Department of Physics and Center for Theoretical Studies of Physical Systems, Clark Atlanta University, Atlanta, Georgia 30314, USA

[2]Department of Physical Chemistry, University of the Basque Country, Leioa, Spain

*zfelfli@cau.edu

*Abstract*

Low-energy electron scattering investigations of the lanthanide atoms (Eu, Nd, Tb and Tm) demonstrate that electron–electron correlation effects and core-polarization interaction are the dominant fundamental many-body effects responsible for negative ion formation. Ramsauer–Townsend minima, shape resonances and binding energies of the resultant anions are identified and extracted from the elastic total cross sections (TCSs) calculated using the complex angular momentum (CAM) method. The large discrepancy between the recently measured electron affinity (EA) of $0.116 \pm 0.013$ eV and the previously measured value of $1.053 \pm 0.025$ eV for Eu is resolved. Also the previously measured EAs for Nd, Tb and Tm are reconciled and new values are extracted from the calculated TCSs. The large EAs found here for these atoms should be useful in negative ion nanocatalysis including methane conversion to methanol without $CO_2$ emission, with significant environmental impact. The powerful CAM method, requiring only a few poles, obtains reliable binding energies of negative ions with no *a priori* knowledge of experimental or other theoretical data and should be applicable to other complex systems for fundamental understanding of their interactions and electron attachment.

**Introduction**
The vast and diverse important applications of the lanthanide atoms, enriched by the presence of the 4f electrons, in various areas of science and technology necessitate the still lacking fundamental understanding of their structure, dynamics and interactions. For instance, $H^-$ conduction in oxyhydrides containing La, Ce, Nd, etc. atoms, important in fast transport and in high-energy storage and conversion devices [1], requires understanding the role of these atoms. Also, some lanthanide atoms are constituents of the vigorously being studied heavy fermion systems [2, 3]. In particular, investigating the structure and the dynamics of atomic Eu through the TCS calculation could lead to a better understanding of the recently discovered $C_{60}$ spin-charging effect[4] caused by such high-spin atoms as Mn and Eu. The important finding in [4] that the Eu atom retains the integrity of its electron spin density when confined within the $C_{60}$ makes it less attractive for building quantum registers. Generally, the calculation of the electron affinities (EAs) of atoms provides a sensitive probe of many electron effects. To date, obtaining accurate and reliable EAs of the lanthanide atoms still continues to present a great challenge for both experiment and theory, see for example [5, 6] and references therein.

Recently, the EA of atomic Eu was measured to be $0.116 \pm 0.013$ eV [5]. This value is in outstanding agreement with the theoretically calculated values using the Regge pole [6] and multi configurational Dirac-Fock-relativistic configuration interaction (MCDF-RCI) [7] methods. Previously, the EA of Eu was measured to be $1.053 \pm 0.025$ eV [8]. Can there be two different EA values for an atom? The answer is emphatically no. To explore the question further, we have adopted the complex angular momentum (CAM) method developed by Connor [9] as implemented in electron scattering [10, 11] and investigated in the electron impact energy range 0.11 eV < E< 4.0 eV the binding energies (BEs) of negative ions formed during the collision of an electron with atomic Eu as Regge resonances. We have extracted the value of 2.63 eV as the EA of Eu. This leads us to conclude that neither the measured EAs of Eu referenced above correspond to the actual EA of Eu. We surmise that both values represent the BEs of excited (metastable) states of the $Eu^-$ anion formed during the collision.

We have also investigated the EA of atomic Nd through the TCS calculation following the measurement [12] which concluded that the EA of Nd was > 1.916 eV. Previously, the EAs of Nd were determined theoretically to be 0.169 eV [13], 0.167 eV [7] and 0.162 eV [6]. The large disagreement between the theoretical EAs on the one hand and the experimental EA on the other for Nd and the problem associated with the EA of atomic Eu referenced above have motivated us to explore at the fundamental level the energy region $0.11 \leq E \leq 4.0$ eV in search of the possible EAs of Nd as well as of atomic Tb and Tm. The choice of the former is dictated to by the experimental suggestion that its EA is > 1.165 eV [12], while for the latter atom the experiment [14] found the EA to be > 1.03 eV indicating in both atoms an order of magnitude stronger interaction between the attaching electron and the neutral atom compared to that found in Eu [5]. In contrast to the experiments [8, 12, 14], the current and existing experiments and theory, see summary in [6], seem to indicate a weaker interaction between the extra electron and the neutral atom. The present investigation therefore also aims *inter alia* to systematically assess the strength of the interaction between the attaching electron and the lanthanide atoms Eu, Nd, Tb and Tm through the control of the electron-electron correlation effects and the core-polarization interaction, thereby probing its dependence on Z and the electronic structure complication of these f-block elements.

The formation, existence and stability of many negative ions are determined mainly by electron–electron correlations and core-polarization interaction [15]. These effects, requiring considerable computational efforts, render the determination of accurate and reliable EAs of complex atoms, such as the lanthanide atoms very challenging for *ab initio* electronic structure theory. This led to the development of the CAM, (also known as the Regge pole) method for application to low-energy electron scattering resulting in electron attachment [10]. The vital importance of the core-polarization interaction in low-energy electron scattering from atoms and molecules was recognized and demonstrated long ago [16-21], including in a recent R-matrix calculation [22]. Indeed, when the core-polarization interaction is accounted for adequately the important characteristic Ramsauer–Townsend (R-T) minima, which manifest the polarization of the atomic core by the scattered electron [23], will appear naturally in the calculated electron elastic scattering TCSs. Also according to the experiment [24] the appearance of a dramatically sharp resonance peak in the low-energy electron-atom scattering TCS is a signature of the formation of a ground state negative ion. This facilitates considerably the identification and extraction of the EAs from the TCSs; it is implemented in the present calculation.

For the calculations carried out here we have selected the CAM method since it has proved effective in calculating the electron elastic scattering TCSs whence we extracted reliable EAs of both the tenuously and weakly bound Ca⁻ and Sr⁻ negative ions, respectively, the structurally complicated Ce atom [11] and the heavy Au atom [25]. In the CAM method is embedded fully the electron–electron correlations, and the core-polarization interaction is incorporated through the well investigated rational function approximation of the Thomas-Fermi (T-F) potential [26, 27] which can be analytically continued into the complex plane. The success of the CAM method in exploring electron attachment in the lanthanide atoms has already been demonstrated [6]. For instance, the recent photodetachment measurements of the EA of the open d- and f-subshell Ce atom [28-30], aided by the multi configurational Dirac-Fock-relativistic configuration interaction calculations [31, 32], are very close to those calculated using the Relativistic energy-consistent small-core lanthanide pseudopotential multireference configuration-interaction (MRCI) [33], and our Regge pole [6] methods. These agreements among the measurements [28-30], the very sophisticated relativistic theoretical calculations [31-33] and the simple-looking nonrelativistic Regge pole theory [6, 11] on the EA of Ce stimulate the compelling question: What is the important fundamental physics common among these diverse theoretical calculations responsible for the excellent agreement with the measurements on the EA of Ce?

The powerful CAM method [9,10,34] is used to calculate the electron elastic TCSs whence is extracted the R-T minima, the shape resonances (SRs) and the binding energies (BEs) of the resultant anions formed during the collision as Regge resonances. Equation (1) below which fully embeds the vital electron-electron correlation effects is used to calculate the TCSs. The potential given by Eq (5) below is used in the calculation of the TCSs. The parameters 'a' and 'b' of the potential (5) determine and control the polarization interaction. For all the calculations performed here the determined optimal value of a=0.2; the 'b' parameter is optimized for each atom as demonstrated under the results. The important quantities also determined here, but not presented in the paper, are the Regge trajectories Re L and Im L (L is the CAM). These probe electron attachment at the fundamental level near threshold since they penetrate the atomic core. The Im L is also used to distinguish between the shape resonances (short-lived resonances) and the stable bound states of the resultant negative ions (long-lived resonances). It must be emphasized that the Regge-pole method requires no *a priori* knowledge of any experimental or other theoretical data as input or guidance; it extracts the EAs from the resonances in the TCSs.

The imaginary part of the complex angular momentum, Im L is an important quantity in our calculations using the CAM method. In Connor [9] and [6] the physical interpretation of Im L is given. It corresponds inversely to the angular life of the complex formed during the collision. A small Im L implies that the system orbits many times before decaying, while a large Im L value denotes a short-lived state. For a true bound state, namely $E < 0$, Im L ≡ 0 and therefore the angular life, 1/[Im L] → ∞, implying that the system can never decay. In our calculations Im L is also used to differentiate between the ground and the excited states of the negative ions formed as resonances during the collisions. The structure of the paper is as follows. Section 2 presents the TCSs, while Sections 3 and 4 deal with the Discussion and Conclusion, respectively. Section 5 presents the brief Theory.

**Results**
Figure 1 displays the electron elastic TCS (a.u.) for atomic Eu versus E (eV) in the electron energy region $0.5 \leq E \leq 4.0$ eV. This typical TCS, see for example the data for Au [25], is characterized by a R-T minimum followed by a shape resonance (SR) and then a second R-T minimum at 2.64 eV. In the region of the second R-T minimum appears the dramatically sharp resonance at an energy which corresponds to the BE of the Eu⁻ negative ion formed during the collision; we identify it with the EA of atomic Eu. The resonance appears when the *'b'* parameter is optimum, corresponding to the value of 0.0432. The appearance of the R-T minima in the TCS, manifests that the polarization interaction is incorporated correctly into the calculation [23].

The great sensitivity to the polarization interaction of the EA of Eu is evident from the values of the *'b'* parameter depicted in the figure, *viz.* when $b = 0.0431$ the resonance disappears. When *'b'* changes by 0.0001 (from 0.0431 to 0.432) the resonance appears, indicating electron attachment resulting in the formation of the negative ion Eu⁻, consistent with the experimental prescription [24]. Also, when the *'b'* parameter changes from 0.0432 to 0.0433 the "resonance" continues to exist, but with its Im L being much larger than that corresponding to $b = 0.0432$. As pointed out by Johnson and Guet [23], when the polarization potential is appropriate the R-T minima will appear naturally.

Sensitivity of the positions of the R-T minima as well as the position of the shape resonance to the parameter *'b'* is not significant. From the figure the EA of Eu is determined to be 2.63 eV. The values of the R-T minima, the shape resonance, and the EA are presented in Table 1, where the current EA is compared with those measured by Refs. [5,8]. It is noted that the recently measured EA [5] and the previously calculated values [6,7] correspond to the BE of an excited state of the Eu⁻ anion and not to the EA of Eu. Also, the shape resonance at 1.43 eV is relatively close to the measured EA value of 1.05 eV [8].

From the above results, we conclude that: 1) neither the measured EAs of Eu [5,8] nor the previously calculated EA values [6,7] correspond to the value of the EA of Eu; and 2) the measured [5] and previously calculated [6,7] EAs of Eu correspond to the second excited state of the Eu⁻ anion. Furthermore, the measured EA [8] of Eu represents the first excited state of the Eu⁻ anion because of its closeness to our value of 1.08 eV. So, both measurements are essentially correct but neither measured the EA of Eu as claimed. These revelations call for both experimental and theoretical verification.

Figure 2 presents the variation of the electron elastic TCS (a.u.) in the energy region $0.02 \leq E \leq 0.4$ eV. The various curves demonstrate the behavior of the TCS with respect to the variation of the *'b'* parameter. Near threshold appears the shape resonance which is followed by the dramatically sharp (long- lived) resonance (red curve) at 0.116 eV. This value, very close to that calculated in [6,7], was also obtained previously by our group [6] and should be compared with the recently measured value of $0.116 \pm 0.013$ eV [5]. The value of Im L of the resonance leads us to conclude that this measured value [5] corresponds to the BE of an excited state of Eu⁻ and not the EA of Eu. Here we also note the sensitivity of the electron attachment to the *'b'* parameter of the polarization potential. When *'b'* changes from 0.0375 to 0.0376 the resonance disappears – no electron attachment results. The other "resonances" simply indicate that the optimal value of the polarization potential has not been realized because their Im L values are larger than that when $b = 0.0375$. The BE obtained from Fig. 2 is also presented in Table 1 for comparison.

In Figure 3 is displayed the electron elastic TCS (a.u.) versus E (eV) for the Nd atom in the electron impact energy region $0 \leq E \leq 3$ eV. As in the case of Eu, the TCS is characterized by a R-T minimum near threshold, followed by a shape resonance and just before the second R-T minimum appears the dramatically sharp resonance when $b = 0.0386$ (purple curve). The resonance at 1.88 eV corresponds to the EA of Nd and is consistent with the experimental value of ~ 1.916 eV [12]. In this case also the sensitivity of the EA to the *'b'* parameter is clearly manifested. Namely, when the value of *'b'* changes from 0.0386 to 0.0385 the resonance disappears, implying that the polarization potential is not appropriate for electron attachment to yield a stable Nd⁻ anion. The other "resonances" that appear as *'b'* increases from its optimal value of 0.0386 have an Im L value that is larger than that corresponding to $b = 0.0386$.

In Figure 4 is plotted the electron elastic TCS (a.u.) versus E (eV) for atomic Nd in the electron impact energy region $0 \leq E \leq 0.5$ eV. As in the case of Eu, the TCS is characterized by a shape resonance near threshold, that is followed by a dramatically sharp resonance at 0.162 eV when the value of 'b' is 0.0340 (orange curve). When 'b' decreases from this optimal value to 0.0339 (purple curve) the resonance disappears, and when 'b' increases beyond its optimal value the Im L value increases as well. The "resonances" appearing beyond $b = 0.0340$ simply indicate that the polarization potential is not appropriate to produce a bound negative ion.

Contrary to Figs. 1-4, in Fig. 5 we simply present the TCSs for the optimum values of a=0.2 and 'b' given in Table 1 for each atom. Figure 5 presents the electron elastic TCSs for the ground state and two excited states in the electron impact energy range $0.01 \leq E \leq 8$ eV. Each of the cross sections is characterized by R-T minima, shape resonances and the dramatically sharp resonances at 0.437 eV, 1.20 eV and 3.04 eV. These correspond to the BEs of the stable anions formed during the collisions as resonances. Of particular interest here is the resonance at 3.04 eV, appearing at the lowest R-T minimum of the TCSs. This new energy is identified with the EA of Tb. Its appearance at the second R-T minimum of the ground state TCS, resembling the behavior of the Au TCS [25], is characteristic of a good negative ion nanocatalyst [35]. It could catalyze a reaction whose vertical detachment energy (VDE) is close to 3.04 within the impact energy range 2.5 eV $\leq E \leq 5$ eV, similarly to the catalysis of $H_2O$ to $H_2O_2$ using the negative ions [36].

The resonance at 1.20 eV, appearing at the second R-T minimum of the TCS for the first excited state, could be useful in catalyzing a reaction whose VDE is around 1.20 eV. This resonance is also of particular importance in that its value is very close to the measured EA of about 1.165 eV [12]. Hence, it is safe to conclude that the measured EA of Tb by [12] corresponds to the BE of the first excited state of the Tb⁻ anion formed during the collision as a resonance and definitely not to the EA. Depending upon the sensitivity of the experiment it could be easier to detect the resonance at 1.20 eV than the EA at 3.04 eV, simply because of the relative magnitudes of their TCSs.

Resonance at 0.437 eV corresponds the BE of the second excited state of the Tb⁻ anion formed during the collision of the electron and the Tb atom. It is also formed at almost the second R-T minimum, which is very shallow in this case. It is noted that the BE of 0.437 eV was incorrectly identified as the EA of Tb in [6]. The curves with the BEs of 3.04 eV and 0.437 eV resemble those of the Au atom [25] and perhaps, Tb could be functionalized similarly.

Each of the curves in Fig. 5 is characterized by two R-T minima which are explained as follows. When the incident electron approaches the Tb atom, it polarizes it and the maximum polarization is represented by the first R-T minimum. Then without conserving angular momentum and energy, the electron becomes trapped by the centrifugal barrier, giving rise to the shape resonance. As the electron leaks out of the barrier, it becomes attached to the atom forming a negative ion, represented by the very thin lines. For both the ground and excited states this occurs at the second R-T minima, representing the maximum polarization of the Tb atom as the electron leaves the atom after the elastic collision. The values of the R-T minima, SRs and the BEs are presented in Table 1.

Because of the appearance of the anionic BEs at the R-T minima of the TCSs of Tb, the Tb anion could be used as a nanocatalyst for various reactions with VDEs of 3.04 eV, 1.20 eV and 0.437 eV. This contrasts with the observation in [37] where the catalysis of $H_2O$ to $H_2O_2$ was effected using Au and Pd nanocatalysts. The significant enhancement of the production of $H_2O_2$ when both atoms were used together [37] was attributed to the fact that the EAs of both the Au and Pd atoms are located at the second R-T minimum of the atomic Au TCS [38]. The VDE of $H_2O$ is located at this minimum as well.

Consequently, the Tb atom through its anion could be utilized to catalyze various reactions at 3.04 eV, 1.20 eV and 0.437 eV as well as a sensor without tuning [39].

Although the TCSs in Fig. 5 appear to be complicated, our Regge pole analysis using Regge trajectories facilitates their understanding considerably. Each component (ground state and excited states TCSs) can be analyzed separately. These results demonstrate that electron attachment resulting in the formation of stable anions should be probed through low-energy electron scattering as was recommended in [24], thereby enabling the extraction of reliable BEs and EAs from the resonances in the elastic TCSs. Furthermore, the appearance of the EA at the R-T minimum of the TCS should be important in the search for negative ion catalysts for the dynamic oxidation of water to peroxide and of methane to methanol without $CO_2$ emission.

As in Fig. 5 here we simply present the TCSs using the optimum values of a=0.2 and 'b' given in Table 1 for atomic Tm. Figure 6 presents the electron elastic TCSs for the ground state and two excited states in the electron impact energy range $0.02 \leq E \leq 10$ eV. Previously, low-energy electron scattering from atomic Tm was investigated through the calculation of the electron elastic TCSs over the energy range $0.001 \leq E \leq 1.0$ eV [40], selected because of the need to understand the very small predicted EA of Tm, see [6]. In [40] it was found that the TCS of Tm was rich in resonances and R-T minima. Additional to the resonance at 0.016 eV found in [6] resonances at 0.104 eV and 0.274 eV were revealed, with the one at 0.274 eV incorrectly identified as the EA of Tm.

In the current energy range $0.02 \leq E \leq 10$ eV the TCSs for Tm generally resemble those of Tb, except that the values of the resonances and the R-T minima are different; consequently, only the essential results will be discussed here. Three dramatically sharp resonances appear in the TCSs for Tm at 0.274 eV, 1.02 eV and 3.36 eV. Interestingly, the resonance at 1.02 eV agrees excellently with that measured in [14] and identified as the EA of Tm. Unfortunately, the measured value does not correspond to the EA of Tm, but to the BE of the first excited state of the Tm⁻ anion formed during the collision as a resonance. The value of 3.36 eV located at the second R-T minimum of the Tm TCS is identified with the EA of Tm and is consistent with that found for the Tb atom. The values for the BEs and the R-T minima are included in Table 1.

We remark here that the Tm atom could be as useful in nanocatalysis and in sensor technology as the Tb atom. Furthermore, except for the ground state TCSs the structure of the cross sections for Tb and Tm are significantly different from each other, manifesting their complex electronic structure configurations. This negates the attempt to generalize the interactions of the lanthanide atoms and reduce them to a single simple formula. The complexity of their electronic structures is what makes them interesting, challenging and important in many scientific and technological fields of applications

**Discussion of Results**
For the four typical lanthanide atoms Eu, Nd, Tb and Tm low-energy electron scattering TCSs have been calculated over the energy range $0.02 \leq E \leq 10$ eV using the CAM method which embeds fully the electron-electron correlation effects. New, never calculated before to our knowledge, characteristic R-T minima, SRs and BEs of the negative ions formed during the collisions have been extracted and compared with the available experimental and other theoretical values. In both atomic Eu and Nd we investigated in low-energy electron scattering the effect of the variation of the polarization interaction on the calculated R-T minima, SRs and the BEs of the resultant negative ions. Figures 1-4 demonstrate the sensitivity of the results to slight variations in the 'b' parameter of the polarization interaction. From the optimized 'b' parameters we calculated the TCSs from which the R-T minima, SRs and the BEs for these atoms were extracted; they are presented in Table 1 where they are compared with available measured and calculated data.

For the Eu atom we have demonstrated that: a) The most recently measured EA [5] which is in outstanding agreement with the theoretical values [6,7], but differing by an order of magnitude from the previously measured EA [8], does not correspond to the EA of Eu; it represents the BE of the second excited state of the Eu⁻ anion and b) The EA measured in [8] corresponds to the BE of the first excited state of the Eu⁻ anion and certainly not to the EA of Eu. We have extracted the value of 2.63 eV as the new EA of Eu. For the Nd atom our calculated EA is very close to that recommended by the measurement [12], but differs significantly from the previously calculated values [6, 7, 13]. Interestingly, for both the Tb and Tm atoms we obtained BEs of their first excited anions that are very close to the measured EAs [12, 14], respectively. These measured EAs were incorrectly identified with the EAs of Tb and Tm. Also found here was the 0.274 eV BE of the second excited state of the Tm⁻ anion, which was incorrectly identified with the EA of Tm in [40]. Our results clarify the disturbing conflicting measurements and sophisticated theoretical calculations on the EAs of the Eu, Nd, Tb and Tm atoms and call for immediate experimental and/or theoretical verification. These results also demonstrate the need for robust theoretical methods, as the CAM method, to map and delineate the complex resonance structures that characterize these lanthanide atoms

As demonstrated by the TCSs of Tb and Tm, the appearance of the bound states of the negative ions, created during the collisions at the R-T minima, provides an excellent environment and mechanism for breaking up molecular bonds in new molecules creation from atoms as well as in catalysis through negative ions. For instance in the oxidation of $H_2O$ to $H_2O_2$ catalyzed by the Au⁻ anion the formation of the Au⁻$(H_2O)_2$ molecular complex provides the mechanism for breaking up the bonds in $H_2O$. The same mechanism has been proposed for breaking up the bonds in $CH_4$ in the oxidation of methane to methanol without $CO_2$ emission when catalyzed by the Au⁻ or other similar anions [41], with far-reaching environmental implications. Importantly, an experimental determination of the R-T minimum in very low-energy F+$H_2$ elastic scattering has been proposed for use to detect virtual states formation [42]. Perhaps, just as in [42] this is the mechanism at play in the creation of the deep R-T minima in the electron elastic scattering TCSs as well, clearly visible in Figs. 5 and 6 (a similar figure for Eu exists but is not shown because of space limitation). Furthermore, the identification that the R-T effect in low-energy elastic scattering and the characteristic photoionization cross sections minima have the same origin [43], allows us to probe deeper into quantum dynamics of chemical reactions using both electron and photon probes.

**Conclusion**
The identification of the major and vital many-body effects responsible for negative ion formation in complex atoms as resonances in low-energy electron elastic scattering and their incorporation in our calculations allowed us to produce unprecedented reliable BEs of the Eu⁻, Nd⁻, Tb⁻ and Tm⁻ anions. This should contribute significantly to the long-overdue fundamental understanding of the structure, dynamics and interactions of the lanthanide atoms, including electron driven processes in general, and produce reliable results. Indeed, resonances represent anion formation through electron attachment [44]; so electron scattering should be used to determine electron affinities of atoms and molecules. Contrary to the finding of a weaker [5] and a stronger [8, 12, 14], interaction between the attaching electron and the neutral lanthanide atoms, our calculations find a much stronger interaction, manifested through the large EAs found for the four lanthanide atoms. We believe that many more lanthanide atoms will likewise reveal new and larger EA values than hitherto believed.

It has been demonstrated here that the popular procedure of photodetaching a negative ion and using the Wigner Threshold Law to determine the EA leads to uncertain or even to incorrect determination of the EAs of structurally rich and complex atoms as the lanthanides. Also existing structure-based theoretical methods fail to guide the measurements, since generally summing a partial wave series with a large number of numerically significant terms leads to no physical insight and to uncertain results as well. Regge poles studies are essential for the fundamental understanding of the behavior of atomic and molecular collision differential cross sections (DCSs) and TCSs. In CAM representation scattering DCSs are described using usually a small number of interfering physically significant amplitudes. And Regge pole trajectories determine the essential structure of the DCSs and TCSs by yielding the R-T minima, bound states and the resonances as well as the scattering amplitudes. Indeed, Regge trajectories provide deep insight into the dynamics of scattering by breaking down the scattering process into its sub-components, manifested through the partial cross sections [6].

**Acknowledgments**
This research is dedicated to Professor Alex Dalgarno, FRS for interest in and unwavering and strong support of our research. Research was supported by the US DOE, Division of Chemical Sciences, Geosciences and Biosciences, Office of Basic Energy Sciences, Office of Energy Research and CAU CFNM, NSF-CREST Program. The computing facilities of the National Energy Research Scientific Computing Center are greatly appreciated.

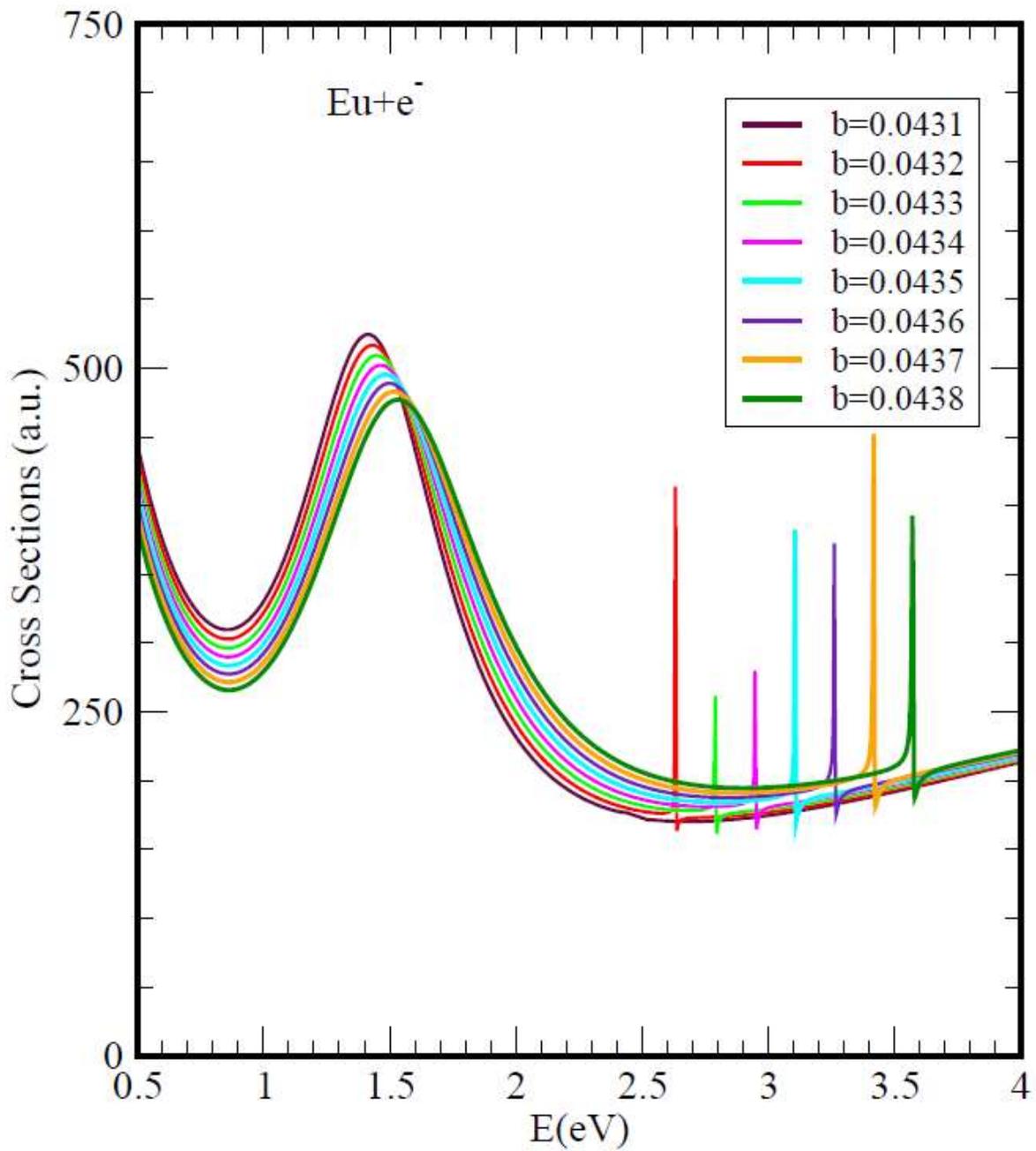

**Figure 1:** Electron elastic scattering TCS (a.u.) versus E (eV) for atomic Eu, demonstrating the sensitivity of the electron affinity (sharp red line) to the 'b' parameter of the polarization potential.

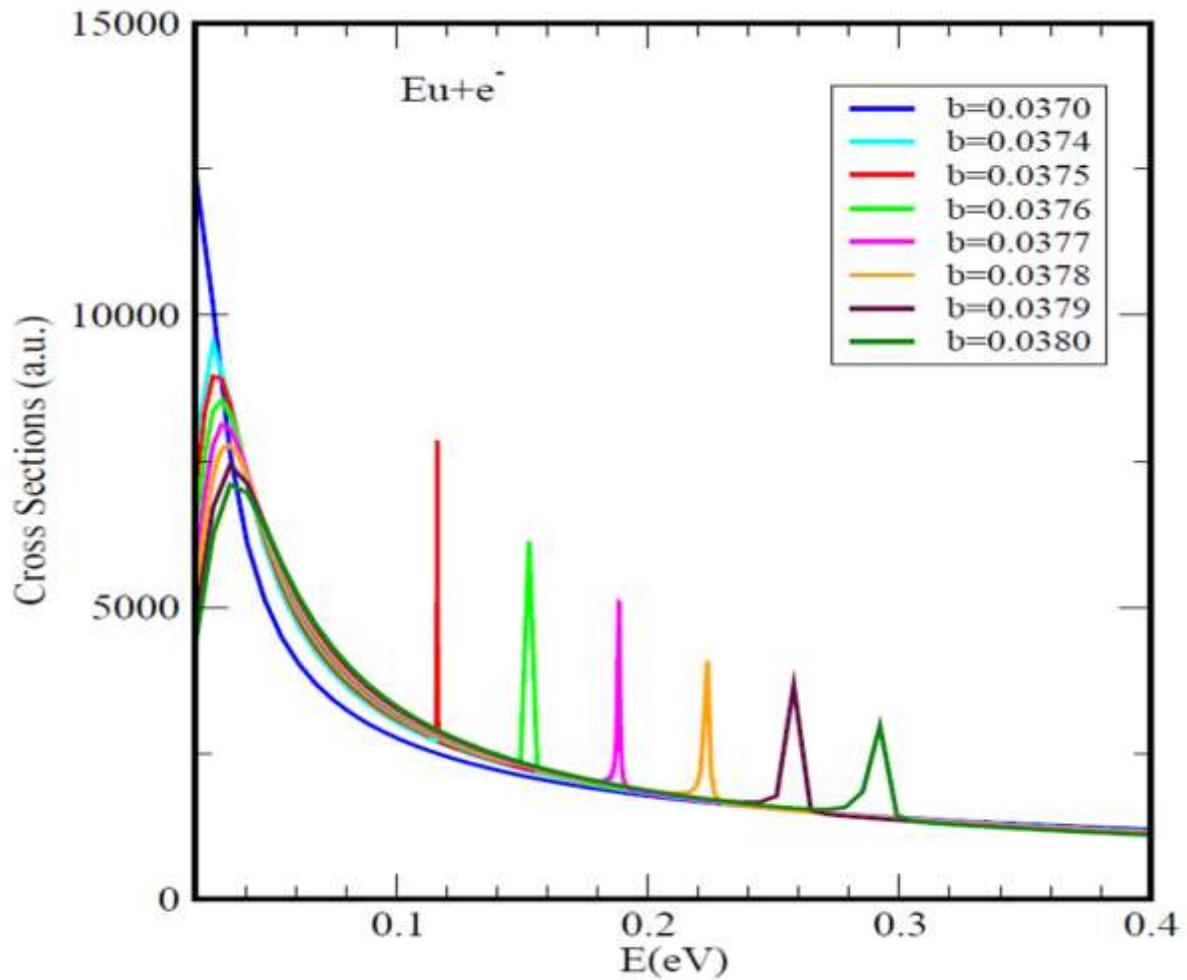

**Figure 2:** Electron elastic scattering TCS (a.u.) versus E (eV) for atomic Eu, demonstrating sensitivity of the binding energy of the excited Eu⁻ anion (sharp red line) to the 'b' parameter of the polarization potential.

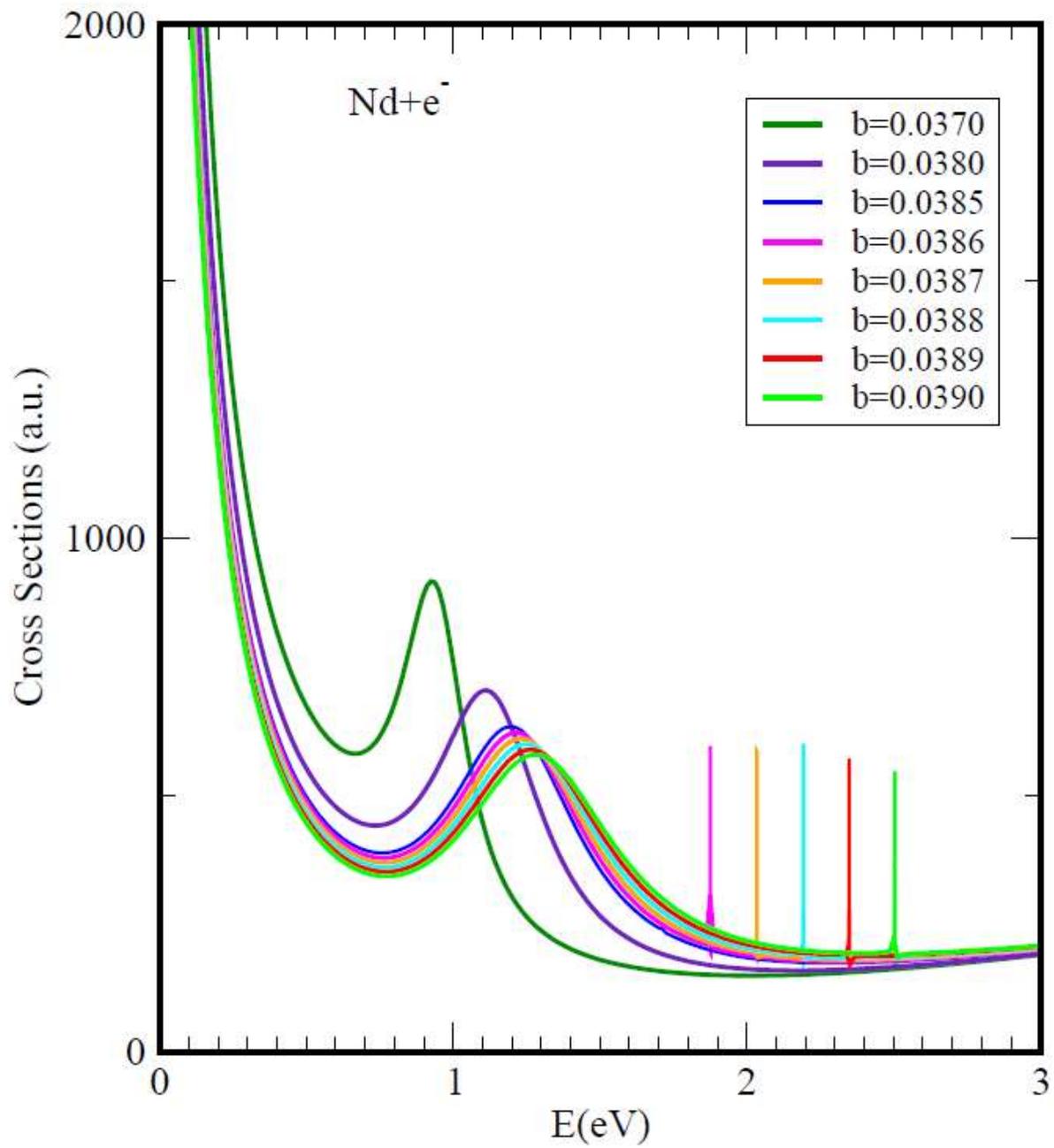

**Figure 3:** Electron elastic scattering TCS (a.u.) versus E (eV) for atomic Nd, demonstrating the sensitivity of the electron affinity (sharp purple line) to the 'b' parameter of the polarization potential.

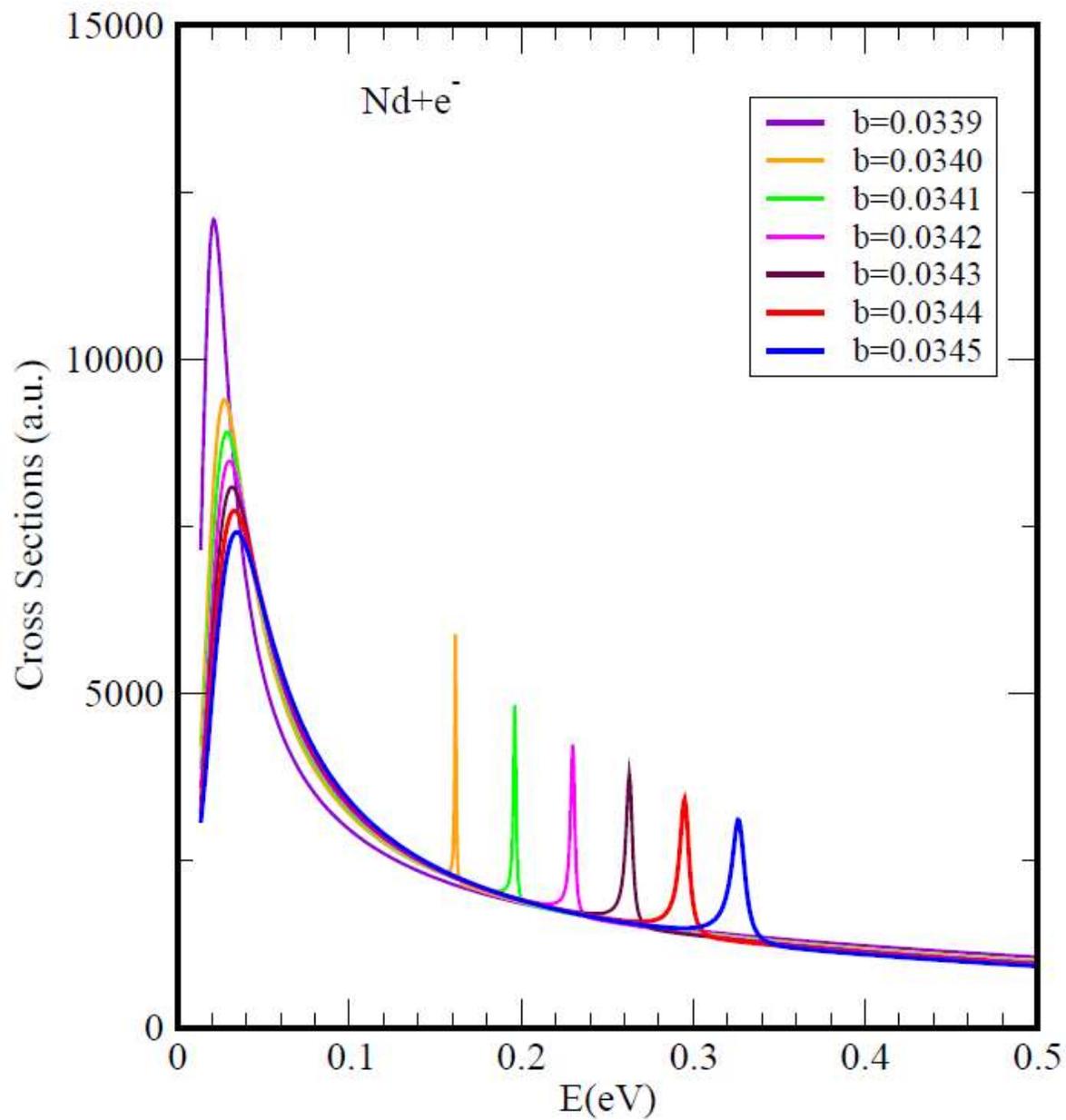

**Figure 4:** Electron elastic scattering TCS (a.u.) versus E (eV) for atomic Nd, demonstrating sensitivity of the binding energy of the excited Nd⁻ anion (sharp orange line) to the 'b' parameter of the polarization potential. Note the sensitivity of the shape resonance as well.

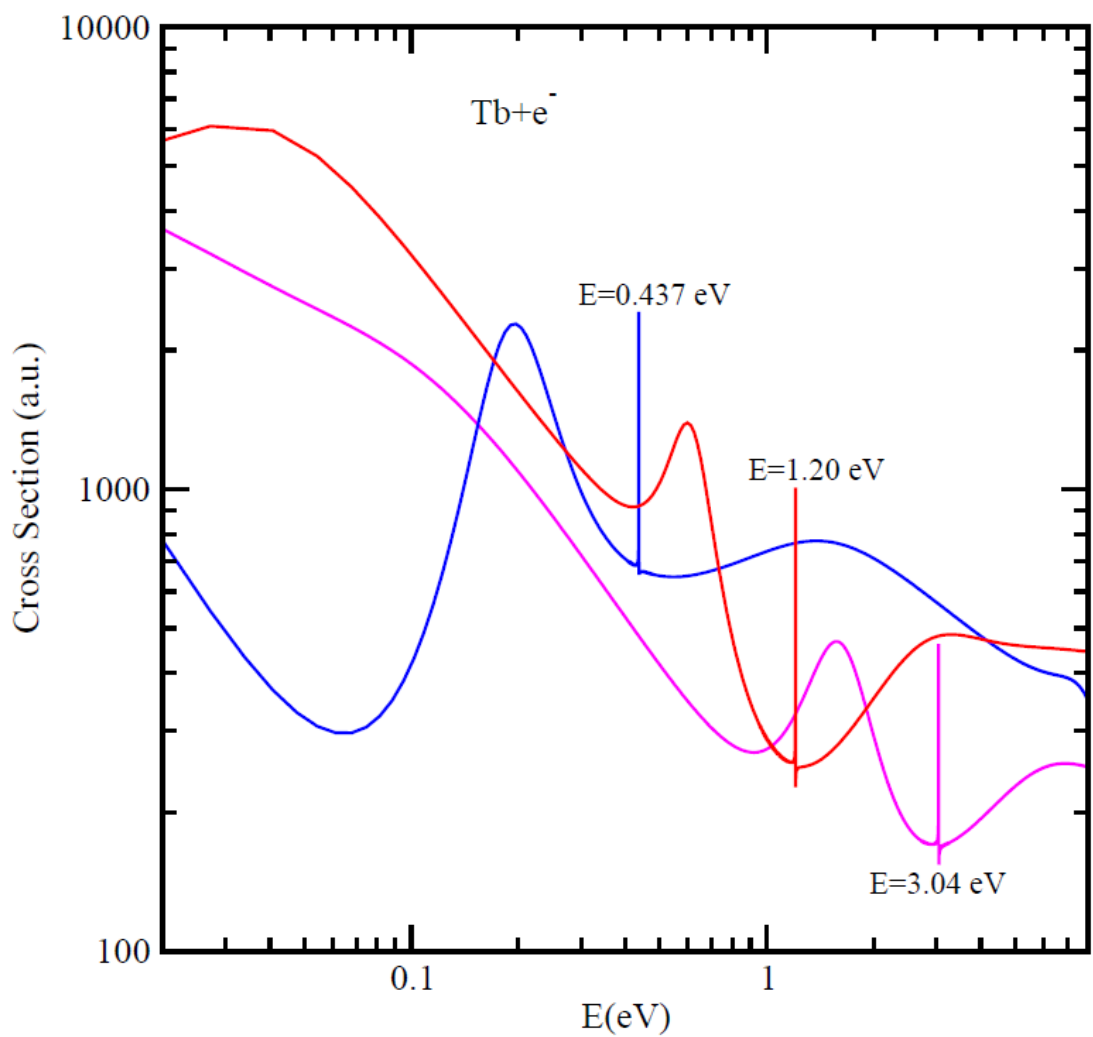

Figure 5. Total cross sections (a.u.) for electron elastic scattering from atomic Tb versus E (eV), are contrasted. The pink, orange and blue curves represent results for the ground, first and second excited states, respectively. All the curves are characterized by R-T minima, shape resonances and dramatically sharp resonance structures corresponding to the formation of Tb⁻ negative ions during the collisions. Note that for the ground-state curve the position of the bound state of the Tb⁻ anion is at the second R-T minimm.

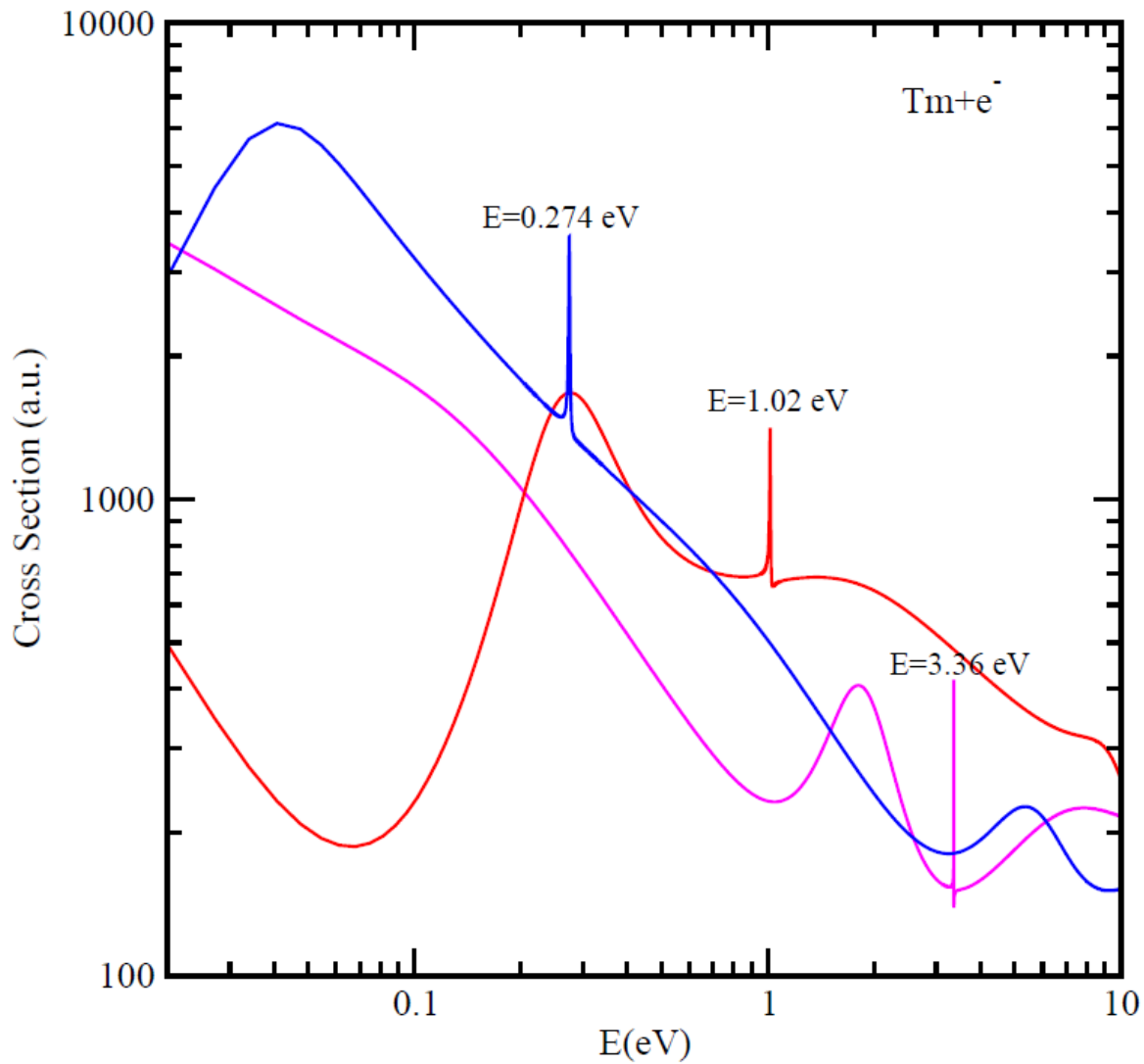

Figure 6. Total cross sections (a.u.) for electron elastic scattering from atomic Tm versus E (eV), are contrasted. The pink, orange and blue curves represent results for the ground, first and second excited states, respectively. All the curves are characterized by R-T minima, shape resonances and dramatically sharp resonance structures corresponding to the formation of Tm⁻ negative ions during the collisions. Note here also that for the ground-state curve the position of the bound state of the Tm⁻ anion is at the second R-T minimm.

**Table 1:** Ramsauer–Townsend (R-T) minima and Shape Resonances (SRs) for the atoms Eu, Nd, Tb, and Tm as well as the Binding Energies (BEs) of their resultant negative ions, all in eV, and optimized 'b' parameters of the polarization potential. RT-1 and RT-2 represent the 1st and 2nd R-T minima, respectively.

| Z | Atom | b | RT-1 | SR | RT-2 | BE | Expt. | Previous Theory |
|---|------|---|------|----|----|----|------|----------------|
| 63 | Eu | 0.0432 | 0.86 | 1.43 | 2.64 | 2.63 | 0.116[5];1.053[8] | 0.116[6]; 0.117[7] |
|   |    |        |      |      |      | 1.08 |                  |                    |
|   |    | 0.0375 | N/A  | 0.029 | N/A | 0.116 |                  |                    |
| 60 | Nd | 0.0386 | 0.76 | 1.21 | 2.32 | 1.88 | >1.916[12]       | 0.169[13];         |
|   |    |        |      |      |      |      | >0.05(AMS) [46]  | 0.167[7]           |
|   |    | 0.0340 | N/A  | 0.027 | N/A | 0.162 |                  | 0.162[6]           |
| 65 | Tb | 0.0463 | 0.92 | 1.57 | 3.05 | 3.04 | >1.165[12]       | 0.085 [7]          |
|   |    | 0.0232 | 0.421 | 0.598 | 1.21 | 1.20 | >0.1(AMS) [46]  |                    |
|   |    | 0.0303 | 0.068 | 0.197 | N/A | 0.437 |                  |                    |
| 69 | Tm | 0.0524 | 1.05 | 1.81 | 3.37 | 3.36 | 1.029[14]        | 0.027–0.136 [47]   |
|   |    | 0.0351 | 0.068 | 0.279 | N/A | 1.02 |                  | 0.032(7) [45]      |
|   |    | 0.0457 | N/A   | 0.041 | 3.26 | 0.274 |                 | 0.022 [7]          |

**Method of Calculation**

The Regge-pole (also known as the CAM) method is appropriate for investigating low-energy electron scattering from the lanthanide atoms since Regge poles, singularities of the S-matrix, rigorously define resonances [48, 49]. The fundamental quantities which appear in the CAM theories are the energy-dependent positions and residues of Regge poles. Plotting Im L(E) versus Re L(E) (L is the complex angular momentum) the well-known and revealing Regge trajectories can be investigated [6]; they probe electron attachment at the fundamental level near threshold since they penetrate the atomic core. Their importance in low energy electron scattering has been demonstrated recently by Thylwe [50]. For the Xe atom the Dirac Relativistic and non-Relativistic calculated Regge trajectories were contrasted and found to yield essentially the same Re L at resonance [50]. This clearly demonstrates the insignificant difference between the Relativistic and non-Relativistic calculations at low scattering energies, corresponding to possible electron attachment, leading to negative ion formation as resonances.

Briefly, in the CAM description of scattering the TCS is given by [10] (atomic units are used throughout):

$$\sigma(E) = 4\pi k^{-2} \int_0^\infty Re[1 - S(\lambda)]\lambda d\lambda - 8\pi^2 k^{-2} \sum_n Im \frac{\lambda_n \rho_n}{1+\exp(-2\pi i \lambda_n)} + I(E) \quad (1)$$

where S is the S-matrix, k = √(2mE), with m being the mass, $\rho_n$ the residue of the S-matrix at the nth pole, $\lambda_n$ and I(E) contains the contributions from the integrals along the imaginary λ-axis; its contribution is negligible [6]. Here we consider the case for which Im $\lambda_n$<<1 so that for constructive addition, $Re\lambda_n \approx 1/2, 3/2, 5/2 \cdots$, yielding $\ell = Re\ L \cong 0,1,2 \cdots$. The significance of Eq. (1) is that a resonance is likely to influence the elastic TCS when its Regge pole position is close to a real integer [10].

In the calculations of the elastic TCSs and the Mulholland partial cross sections we use the well investigated rational function approximation of the Thomas-Fermi (T-F) potential [26, 27]

$$U(r) = \frac{-Z}{r(1+aZ^{1/3}r)(1+bZ^{2/3}r^2)} \tag{2}$$

where $Z$ is the nuclear charge and $a$ and $b$ are parameters. For small r, the potential describes the Coulomb attraction between an electron and a nucleus, $U(r) \sim -Z/r$, while at large distances it mimics the polarization potential, $U(r) \sim -1/(abr^4)$ and accounts properly for the vital core-polarization interaction at low energies. Here the effective potential

$$V(r) = U(r) + L(L+1)/2r^2 \tag{3}$$

is a continuous function of the variables $r$ and L. When the TCS as a function of 'b' has a resonance [10] corresponding to the formation of a stable bound negative ion, this resonance is longest lived for ground state collisions and fixes the optimal value of 'b' in Eq. (5). For all the cases considered here the optimal value of $a = 0.2$.

For the numerical evaluation of the TCSs and the Mulholland partial cross sections, we solve the Schrödinger equation for complex values of L and real, positive values of E

$$\psi'' + 2\left(E - \frac{L(L+1)}{2r^2} - U(r)\right)\psi = 0, \tag{4}$$

with the boundary conditions:
$\psi(0) = 0$,
$\psi(r) \sim \exp(+i\sqrt{2Er}), r \to \infty$ \hfill (5)

Equation (8) defines a bound state when k ≡ √(2E) is purely imaginary positive. The S-matrix, S(L, k), poles positions and residues of Eq. (7) are calculated following a method similar to that of Burke and Tate [51]. In the method the two linearly independent solutions of the Schrödinger equation are evaluated as Bessel functions of complex order and the S-matrix, which is defined by the asymptotic boundary condition of the solution of the Schrödinger equation, is thus evaluated. Further details of the calculation may be found in [34, 51].